\documentstyle[12pt]{article}
\begin{document}
\title{{\bf OBSERVATIONAL CONSEQUENCES
\\ OF MANY-WORLDS QUANTUM THEORY}
\thanks{Alberta-Thy-04-99, quant-th/9904004}}
\author{
Don N. Page
\thanks{Internet address:
don@phys.ualberta.ca}
\\
CIAR Cosmology Program, Institute for Theoretical Physics\\
Department of Physics, University of Alberta\\
Edmonton, Alberta, Canada T6G 2J1
}
\date{(1999 May 3)}
\maketitle
\large

\begin{abstract}
\baselineskip 16 pt

	Contrary to an oft-made claim,
there can be observational distinctions
(say for the expansion of the universe
or the cosmological constant)
between ``single-history'' quantum theories
and ``many-worlds'' quantum theories.
The distinctions occur
when the number of observers
is not uniquely predicted by the theory.
In single-history theories, each history is weighted
simply by its quantum-mechanical probability,
but in many-worlds theories
in which random observations are considered,
there should also be the weighting by the numbers
or amounts of observations occurring in each history. 

\end{abstract}
\normalsize
\baselineskip 16 pt
\newpage

	Quantum mechanics is so mysterious that
its precise content or interpretation
is not agreed upon even
by leading physicists.
Although the number of versions or interpretations
of quantum mechanics is huge,
here I wish to focus
upon two main classes of interpretations,
which I shall call ``single-history'' versions
and ``many-worlds'' versions,
and show how they might be distinguished
observationally.
(Similar observational distinctions
can be made between analogous
``single-history'' and ``many-worlds'' versions
of classical physics,
but since we know that the universe
is quantum, here I shall focus
on quantum theories.)

	In single-history versions,
the quantum formalism gives probabilities for
various alternative sequences of events,
but only one choice among
the possible alternatives is assumed to occur in actuality.
For example, a wavefunction that gives nonzero
amplitudes for many different alternative events
may be assumed to undergo
a sequence of collapses to
give a single sequence of actually occurring events,
which may be considered to be a unique history.

	On the other hand,
the many-worlds versions
began with Everett's relative-state formalism
\cite{mw}
in which the wavefunction never collapses.
In a suitable basis
each component of the wavefunction
may be considered to be a different ``world,''
leading to this interpretation's being
labeled the ``many-worlds'' interpretation.

	The consistent or decohering histories
formulation of quantum mechanics
\cite{ch}
does not by itself imply whether only
a single coarse-grained history actually
occurs, or whether many do,
and the probabilities of histories
that it gives do not depend on whether
only one, or instead many, of the histories
are actual rather than merely possible.
However, I am considering
probabilities for observations rather than
merely probabilities for histories,
so the consistent or decohering histories
formalism needs to be extended in order
to calculate these probabilities of interest here.
The extension then depends on whether
many, or only one, of the histories are actual.

	It is often claimed that there is no observational
distinction between many-world and single-history
versions of a quantum theory
\cite{noob},
but here I shall refute that.

	In processes with fixed observers
that remember their observations,
it does seem to be true that
there is generally no distinction
that a single observer can make
between single-history
and many-worlds quantum theories
that are otherwise identical.
This is because
then the measure for each observation
in a many-world theory
is proportional to the probability
of that observation in
the corresponding single-history theory.
This result depends upon the lack of
interference between ``worlds''
in which different observations are made,
which is assured if the memory records
of the different observations are orthogonal.	

	To circumvent
this no-observable-distinction result,
David Deutsch
\cite{Deu}
has proposed an experiment
in which an observer ``splits'' into two
copies which make different observations
and remember the fact of observation,
but not the distinct observations themselves,
and so can in principle be
rejoined coherently back into a single copy.
However, doing this in practice appears
to be technologically extremely challenging.

	On the other hand,
what I wish to demonstrate here is that
if different ``worlds''
do not have the same number of observers,
then the measures for observations
in the many-worlds theory can be
different from being merely proportional
to the probabilities in the corresponding
single-history theory.
Then what an observer would be typically
expected to observe in the two theories
can be distinct.

	Consider a theory of quantum cosmology
that gives a quantum state for the universe
in which there are different ``worlds''
with greatly different numbers of observers.
For calculating how typical various observations are,
in a single-history theory one should weight
the ``worlds'' purely by how probable they are,
but in a many-worlds theory, one should weight
the ``worlds'' not only by their quantum mechanical
measures (the analogue in a deterministic
many-worlds theory of the probabilities
in an indeterministic single-histories theory),
but also by how much observation
occurs within each ``world.''
This distinction leads to different predictions
as to which observations would be typical
within the two types of theories.

	As a grossly oversimplified illustration,
consider the example in which a quantum cosmology
theory gave a quantum state (before any possible
collapse) that had one ``world''
with observers, and a second one with none.
Suppose that the first ``world'' had a measure
of 0.0000000001
and the second one had a measure of 0.9999999999.

	In the single-history version of this theory,
these two normalized measures would be the
probabilities for the two ``worlds,''
so the probability would be extremely low
that this theory led to any observers.
A non-null observation would thus have such a low
likelihood within this single-history theory
that it would be strong evidence against
this theory.

	On the other hand,
in the many-worlds version of this theory,
both ``worlds'' would exist,
with the measures indicating something
like the ``amount'' by which they exist.
But since the observations that occur
in the first ``world'' definitely exist
within this many-worlds theory
as realities and not just as possibilities,
the existence of an observation is
not evidence against this many-worlds theory.

	To put it another way,
for considering observations
within a many-worlds theory,
one must multiply the measure
for each world by a measure
for the observations within that world.
(Crudely, one may use the number of observations
within the world, though in a final theory
I would expect a refinement,
so that, for example, a human's observation
is weighted more heavily than an ant's).
When one does this for the example above,
the first ``world'' makes up the entirety
of the weighting in the many-worlds theory,
even though in the single-history theory
that ``world'' has an extremely low probability
and would be quite unexpected.

	Now consider a second toy theory
in which there are two ``worlds''
that both have observers,
but their numbers and observations differ.
For example, let World A last just barely long enough
for it to have $10^{10}$ observers,
all during the recontracting stage fairly near a big crunch,
and let World B last much longer
than the age range at which observers occur
and have $10^{90}$ observers when the universe is expanding.
Suppose World A has measure almost unity
and World B has measure $10^{-30}$.

	In the single-history version of this theory,
these (normalized) measures are probabilities,
so with near certainty,
we can deduce that we should
be in World A and see a contracting universe in this theory.  
Our actual observation of an expanding universe
would then be strong evidence against this
single-history theory.

	On the other hand,
in the many-worlds version of this theory,
all of the observations actually exist.
To calculate which observations are typical,
one needs the measures for the observations themselves.
Presumably these are given by the expectation values
of certain operators associated with the corresponding
observations
\cite{SQM}.
Crudely one might suppose the
total for all the observations within one ``world''
is roughly proportional to the number
of observers within that ``world,''
multiplied by the measure for the ``world.''
At this level of approximation,
the total measure for the observations
in the many-worlds version of this second toy model is thus
$10^{10}$ for World A and $10^{60}$ for World B.
Therefore, an observation chosen at random
in this many-worlds theory is $10^{50}$
more likely to be from World B,
with the universe observed to be expanding,
than from World A with the universe seen to
be contracting.
Our actual observation of an expanding universe
would then be consistent with this theory.

	Thus in this second toy cosmological model,
we can reject its single-history version
because of the low probability it gives,
not for our existence this time, but for
whether we see the universe expanding.
In this way observations can in principle be used
to distinguish between many-worlds
and single-history quantum theories.

	In these examples,
the statistical predictions of what a random observer
should be expected to observe would be
the same for a many-worlds theory
and for the corresponding single-history theory
if the latter had its quantum-mechanical probability
for each history also weighted by
the number of observers in that history,
but I am assuming that this is not the case.
Note that one could still get observable distinctions
even if the single-history theory
had a sequence of wavefunction collapses,
each of which had the weighting by the number
of observers in each branch at the time of the collapse,
but I shall not consider further this possibility either.

	There is the challenge that at present
we apparently do not know enough
about the quantum state of the universe
to say with certainty whether our observations favor
a many-worlds theory or a single-history theory.
Nevertheless, I can summarize some highly
speculative evidence that gives
a preliminary suggestion that
a many-histories theory might be
observationally favored.

	This evidence starts with the Hartle-Hawking
`no-boundary' proposal for the quantum state
of the universe
\cite{nb},
which of course is quite speculative
but seems to me to be the most elegant
sketch so far of a proposal
(certainly not technically complete at present)
for the quantum state of the universe.
Under certain unproven assumptions and approximations,
in a homogeneous, isotropic three-sphere
minisuperspace toy model
with a single massive inflaton scalar field,
the no-boundary proposal leads
in the semiclassical regime to a set of
``worlds'' or macroscopic classical spacetimes
that are Friedmann-Robertson-Walker universes
with various amounts of inflation
and hence various total lifetimes and maximum sizes,
and with measure approximately proportional to
$e^{\pi a_0^2}$, where $a_0$ is the radius of the
Euclidean four-dimensional hemisphere
where the solution nucleates
\cite{nb}.

	This nucleating radius $a_0$ is inversely
proportional to the initial value of the inflaton scalar field,
multiplied by its mass $m$ in Planck units,
whereas the growth factor during inflation,
and the lifetime of the resulting
Friedmann-Robertson-Walker universe,
go exponentially with the square of
the initial value of the inflaton scalar field.
Therefore, if one works out the quantum measure
in terms of the volume of the universe
at the end of inflation
(say $V$ in Planck units),
one finds that at the tree level
it is very roughly proportional
to $\exp{[(4.5\pi/m^2)/(\ln{m^3 V}+1.5\ln{\ln{m^3 V}})]}$
for large values of $m^3 V$ (the universe volume
in units of the cube of the reduced Compton wavelength
of the inflaton scalar field).

	Since the inflaton mass $m$ is very small
in Planck units, say roughly $10^{-6}$
\cite{Lin90},
the factor of $4.5\pi/m^2$ is very large,
say roughly $10^{13}$,
and one gets an utterly enormous
exponential peak in the measure
at relatively small values of $m^3 V$.
(There is a cutoff in $m^3 V$ at a value of order
unity, below which there is no inflationary solution
\cite{GR},
so the measure distribution does not actually
have a divergence.)
The expression above for the measure rapidly decreases
with increasing volume and then
flattens out to become asymptotically constant
when $m^3 V$ gets large in
comparison with $\exp{(4.5\pi/m^2)}$.

	If one takes at face value
the expression above for the measure
for all values of $m^3 V$ above its lower cutoff
(at some number of order unity),
then although the measure has an
utterly enormous exponential peak
at small values of $m^3 V$,
this is in turn overwhelmed
by the divergence one gets when one integrates
the measure (actually a measure density)
to infinite values of $m^3 V$.
Then the total measure would be completely dominated
by universes with arbitrarily large amounts of inflation.
This means that with unit normalized probability,
our universe would be arbitrarily large and arbitrarily flat
when one ignores density fluctuations
from corrections to the homogeneous isotropic
minisuperspace model
\cite{HP}. 

	However, the expression above for the measure
is purely at the tree-level or zero-loop approximation,
ignoring prefactors that are expected to distort
the measure distribution significantly
for $m^3 V$ large in comparison
with $\exp{(4.5\pi/m^2)}$,
because these enormous universes are generated
by inflation that starts with the inflaton
potential exceeding the Planck density,
where one cannot trust the tree-level approximation
or any other approximation we have at present.

	If the correct quantum measure distribution
diverges when one integrates to infinity
the spatial volume shortly after the end of inflation,
then the universe is most probably arbitrarily large
and very near the critical density (spatially very flat),
whether a many-worlds or a single-history
quantum theory is correct,
and so our observation of a universe
near the critical density would not distinguish
between the two possibilities.

	However, if the correct quantum measure
density is cut off or damped for large initial values
of the inflaton energy density
so that one does not get arbitrarily large
universes with certainty,
then the enormous exponential peak
in the distribution at small universes
is likely to dominate and
(in a single-history version in which
the quantum state collapses to a single
macroscopic Friedmann-Robertson-Walker universe)
make the universe most probably
have only a small amount of inflation
and a very short lifetime,
not sufficient to produce observers,
like the first world in the first example above.
If one said that somehow the quantum state
collapsed to a Friedmann-Robertson-Walker universe
that gets large enough for observers,
then the most probable universe history
under this requirement
would be one that lasts just barely long enough
for observers before the final big crunch.
In this case the observers would most likely
exist only near the end of the universe,
when it is recollapsing,
like World A in the second example above,
which is contrary to our observations of
an expanding universe.
Thus a single-history version of this
theory with the quantum measure cut off
to produce a normalizable probability
distribution would most likely be refuted
by our observations of an expanding universe.

	On the other hand, if one took
a many-worlds version of this quantum
cosmology theory, one would have to weight the
``worlds'' (classical universes)
by something like the number of observers
within them.  One would expect this number
to be proportional to the volume of space
at the time and other conditions
when observers can exist
(other factors being equal)
\cite{vol}.
Therefore,
in the many-worlds version one would
multiply the quantum measure given above
for the ``worlds''
(the ``bare'' probability distribution for
universe configurations
\cite{Page})
by something like $V$ to get the measure for observations
(the ``observational'' probability distribution
\cite{Page}).

	The result,
$V\!\exp{[(4.5\pi/m^2)/(\ln{m^3 V}\!+\!1.5\ln{\ln{m^3 V}})]}$,
is then sufficiently rapidly rising with large $m^3 V$
that the part with large $m^3 V$, even if cut off
at $m^3 V$ of order $\exp{(4.5\pi/m^2)}$,
dominates over the exponentially large
peak near the minimum value of $m^3 V$.
There is thus enough space for the no-boundary
proposal to be consistent with our observations
of a large and expanding universe \cite{Page},
but this argument implicitly assumed
a many-worlds version of the no-boundary proposal.
A similar assumption had been made earlier
in the broader context of eternal stochastic inflation
\cite{vol}.
In a single-history version, it seems plausible
that the Hartle-Hawking `no-boundary'
quantum state may collapse
with nearly unit probability to a classical
universe configuration that only lasts
of the order of the Compton wavelength
of the inflaton scalar field,
presumably far too short to be consistent with
our observations.

	This suggestive evidence against
a single-history quantum cosmology theory
is of course not yet conclusive,
since we do not yet know what the quantum
state of the universe is.
Indeed, the `tunneling' wavefunction proposals
of Vilenkin, Linde, and others
\cite{tun}
predict that the ``bare'' quantum measure
for small universes is exponentially suppressed,
rather than enhanced as discussed above for
the Hartle-Hawking `no-boundary' proposal.
The `tunneling' proposals would thus apparently
be consistent with our observations
whether one used a many-worlds version
or a single-history version.
But the possibility is open that
increased theoretical understanding
of quantum cosmology may lead us to favor
a quantum theory, such as the `no-boundary'
one may turn out to be when it is better understood,
that is consistent with our observations
only in its many-worlds version rather
than in its single-history version.

	Another tentative piece of observational
evidence in favor of many-worlds quantum theory
is a comparison with the calculation
\cite{MSW}
of likely values of the cosmological constant.
If the assumptions of that paper are correct,
and if the ``subuniverses'' used there
are the ``worlds'' used here
(``terms in the state vector'' \cite{MSW})
rather than different spacetime regions
within one ``world'' (``local bangs'' \cite{MSW}),
then our observational evidence of
the cosmological constant is consistent
with many-worlds quantum theory
but not with single-history quantum theory.
However, we need a better understanding
of physics to know whether the assumptions
are correct (such as the assumption that
``the cosmological constant takes a variety
of values in different `subuniverses' '' \cite{MSW}).

	Therefore, it may turn out,
when we better understand
fundamental physics and quantum cosmology,
that the observational evidence
of the expansion of the universe
and of the cosmological constant
may lead us to favor
many-world quantum theories
over single-history quantum theories.

	I am grateful for very helpful discussions
with Meher Antia, Jerry Finkelstein, Jim Hartle,
and Jacques Mallah.
This research was supported in part by
the Natural Sciences and Engineering Research
Council of Canada.



\baselineskip 4pt

\begin{thebibliography}{99}

\bibitem{mw}
H. Everett, III, Rev.\ Mod.\ Phys.\ {\bf 29}, 454 (1957);
B. S. DeWitt and N. Graham, eds.,
{\em The Many-Worlds Interpretation
of Quantum Mechanics}
(Princeton University Press, Princeton, 1973).

\bibitem{ch}
R. B. Griffiths, J.\ Stat.\ Phys.\ {\bf 36}, 219 (1984);
Am.\ J.\ Phys.\ {\bf 55}, 11 (1987);
R. Omn\`es, J.\ Stat.\ Phys.\ {\bf 53}, 893 (1988);
{\bf 53}, 933 (1988); {\bf 53}, 957 (1988);
{\bf 57}, 357 (1989); {\bf 62}, 841 (1991);
M. Gell-Mann and J. B. Hartle, in
{\em Proceedings of the Third International Symposium
on the Foundations of Quantum Mechanics
in the Light of New Technology}, edited by
S. Kobayashi, H. Ezawa, Y. Murayama, and S. Nomura
(Physical Society of Japan, Tokyo, 1990), p.~321;
Phys.\ Rev.\ D{\bf 47}, 3345 (1993).

\bibitem{noob}
B. S. DeWitt, Physics Today {\bf 23}, 30 (Sept. 1970);
R. Omn\`{e}s,
{\em The Interpretation of Quantum Mechanics},
pp. 327, 345 (Princeton University Press, Princeton, 1994);
K. W. Ford and J. A. Wheeler,
{\em Geons, Black Holes, and Quantum Foam:
A Life in Physics}, p. 270 (W. W. Norton, New York, 1998).

\bibitem{Deu}
D. Deutsch, Int.\ J.\ Theor.\ Phys.\ {\bf 24}, 1 (1985).

\bibitem{SQM}
D. N. Page, "Sensible Quantum Mechanics:  Are Only
Perceptions Probabilistic?" (University of Alberta report
Alberta-Thy-05-95, June 7, 1995), quant-ph/9506010;
Int.\ J.\ Mod.\ Phys.\ {\bf D5}, 583 (1996).

\bibitem{nb}
S. W. Hawking, in {\em Astrophysical Cosmology:
Proceedings of the Study Week on Cosmology and Fundamental Physics},
edited by H. A. Br\"{u}ck, G. V. Coyne and M. S. Longair
(Pontificiae Academiae Scientiarum Scripta Varia,
Vatican, 1982);
J. B. Hartle and S. W. Hawking, 
Phys.\ Rev.\ {\bf D28}, 2960 (1983);
S. W. Hawking, Nucl.\ Phys.\ {\bf B239}, 257 (1984).

\bibitem{Lin90}
A. Linde, {\em Particle Physics and Inflationary Cosmology}
(Harwood, Chur, 1990).

\bibitem{GR}
L. P. Grishchuk and L. V. Rozhanskii,
Phys.\ Lett.\ {\bf 208B}, 369 {1988}.

\bibitem{HP}
S. W. Hawking and D. N. Page,
Nucl.\ Phys.\ {\bf B264}, 185 (1986).

\bibitem{vol}
A. D. Linde, Mod.\ Phys.\ Lett.\ {\bf 1A}, 81 (1986);
Phys.\ Lett.\ {\bf 175B}, 395 (1986);
Physica Scripta {\bf T15}, 169 (1987);
M. Aryal and A. Vilenkin,
Phys.\ Lett.\ {\bf 199B}, 351 (1987);
M. Miji\'{c}, Phys.\ Rev.\ {\bf D42}, 2469 (1990);
A. D. Linde, D. A. Linde, and A. Mezhlumian,
Phys.\ Rev.\ {\bf D49}, 1783 (1994);
A. Vilenkin,
Phys.\ Rev.\ Lett.\ {\bf 74}, 846 (1995);
Phys.\ Rev.\ {\bf D52}, 3365 (1995).

\bibitem{Page}
D. N. Page, Phys.\ Rev.\ {\bf D56}, 2065 (1997).

\bibitem{tun}
A. Vilenkin, Phys.\ Lett.\ {\bf 117B}, 25 (1982);
Phys.\ Rev.\ {\bf D27}, 2848 (1983);
{\bf D30}, 509 (1984);
Nucl.\ Phys.\ {\bf B252}, 141 (1985);
Phys.\ Rev.\ {\bf D33}, 3560 (1986);
{\bf D37}, 888 (1988);
{\bf D39}, 1116 (1989);
{\bf D50}, 2581 (1994);
A. D. Linde, Zh.\ Eksp.\ Teor.\ Fiz.\ {\bf  87}, 369 (1984);
Lett.\ Nuovo Cimento {\bf 39}, 401 (1984);
Ya. B. Zel'dovich and A. A. Starobinsky,
Pis'ma Astron.\ Zh.\ {\bf 10}, 323 (1984).

\bibitem{MSW}
H. Martel, P. R. Shapiro, and S. Weinberg,
Ap.\ J.\ {\bf 492}, 29 (1998).

\end{thebibliography}
\end{document}